\documentclass[aps,twocolumn, groupedaddress]{revtex4}

\usepackage{graphicx}% Include figure files
\usepackage{bm}% bold math

\begin{document}
\title{Gas phase sorting of nanoparticles}
\author{Hendrik Ulbricht, Martin Berninger, Sarayut Deachapunya,
Andr\'{e} Stefanov and Markus Arndt}

\affiliation{Quantum Optics, Quantum Nanophysics and Quantum
Information, Faculty of Physics, University of Vienna,
Boltzmanngasse 5, A - 1090 Vienna, Austria}

\date{\today}
\begin{abstract}
We discuss Stark deflectometry of micro-modulated molecular beams
for the enrichment of biomolecular isomers as well as single-wall
carbon nanotubes and we demonstrate the working principle of this
idea with fullerenes. The sorting is based on the species-dependent
polarizability-to-mass ratio $\alpha/m$. The
 device is compatible with a high molecular throughput,
and the spatial micro-modulation of the beam permits to obtain a
fine spatial resolution and a high sorting sensitivity.
\end{abstract}

%\pacs{33.15.Kr, 39.90.+d, 61.46.Fg, 61.48.+c, 81.07.De, 87.14.-g}
\maketitle

Sorting of nanoparticles is essential for many future
nanotechnologies. Nanoparticles can generally be sorted by their
different physical or chemical properties. The objective is to
prepare or enrich a particular species with a distinct property. In
the case of carbon nanotubes the sorting of species with different
metallicity is essential for many applications such as the
realization of field effect transistors, light emitting diodes or
conducting wires~\cite{Avouris2001}. Here sorting can for instance
be achieved by exploiting the tube's dielectric properties in a
liquid environment~\cite{Krupke2003a}. Also chemical methods for the
selection and separation of carbon nanotubes are currently beeing
investigated~\cite{Arnold2006a}.

Complementary to these efforts also the manipulation of large
clusters and molecules in the gas phase has attracted a growing
interest over recent years, in particular with applications in
molecule
metrology~\cite{Bonin1997a,Compagnon2001a,Berninger2007a,Deachapunya2007a}.
Since many nanoparticles, among them biomolecules or carbon
nanotubes, exist in various different isomers and conformations, it
is intriguing to investigate sorting methods in the gas phase which
select the particles according to their polarizability-to-mass ratio
$\alpha/m$ instead of their mass alone.

A large number of classical deflection experiments have been
performed in the past (for a review see~\cite{Bonin1997a}) which
employ the deflection of a well-collimated neutral beam in the
presence of a static transverse inhomogeneous electric field. In
this arrangement, one can usually chose between a wide molecular ray
of high flux or a narrow beam with a lower total signal whose
lateral shift can be determined with higher precision.

We here present a method for sorting nanoparticle beams which
combines high transmission {\em and} high resolution. This can be
achieved by imprinting a very fine spatial modulation onto the
molecular beam.

\begin{figure}
  \includegraphics[width=\columnwidth]{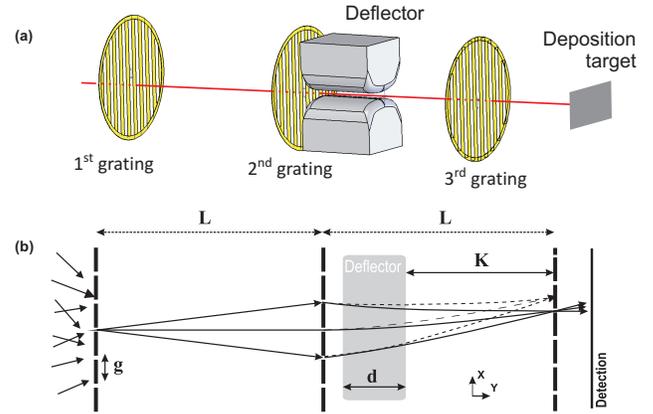}\\
  \caption{(a) Three grating deflection setup. The third grating can be shifted to scan the
  nanoparticle fringe pattern. Particles with different $\alpha/m$ are
   separated by their different deflection shifts in the electrode
   field as identified in (b). The grating position can be set to preferentially
   transmit one species while blocking the others. After the sorting
   the molecules may be deposited on a target or detected by
   ionization.
   }\label{setup}
\end{figure}

Our starting point is a three-grating matter-wave interferometer
which we already described before~\cite{Brezger2003a}. As shown in
Fig.~\ref{setup}, it is composed of three micro-machined gratings,
which prepare, sort and detect the molecules. The combination of the
first two gratings modulates the particle flux such as to generate a
periodic particle density pattern in the plane of the third grating.
All gratings and also the molecular micro-modulation have identical
periods. The density pattern or contrast function can therefore be
revealed by scanning the third grating while counting all
transmitted molecules, as shown in Fig.~2.

Our device is usually operated in a quantum mode, with molecular
masses and velocities chosen such as to reveal fundamental quantum
phenomena related to matter-wave diffraction~\cite{Arndt2005b}.

However, the same device can also be used in a  Moir\'{e} or shadow
mode~\cite{Oberthaler1996a}, where the molecules can be approximated
by classical particles. This applies in particular to fast and very
massive molecules where quantum wave effects may be too small to be
observed.

Our setup then still combines a fine spatial micro-modulation with
much relaxed requirements on the collimation of the beam. This
allows us to increase the spatial resolution in any
beam-displacement measurements by several orders of magnitude over
earlier experiments without micro-imprint.

A beam-displacement may for instance be caused by an inhomogeneous
electric field acting on the polarizability of the particle. In our
experiment of Fig.~\ref{setup}, a pair of electrodes close to the
second grating generates a constant force field $F_{x} = \alpha
(\mathbf{E\nabla} )E_{x}$, which shifts the molecular fringe pattern
along the x-axis by
\begin{equation}\label{shift}
    \Delta s_x \propto (\alpha/m)\cdot (\mathbf{E}
\mathbf{\nabla})E_{x}/v_y^{2}.
\end{equation}
Here v$_y$ is the beam velocity in the forward direction. Deflection
measurements then allow to derive precise values for the
polarizability of the molecules, as recently
demonstrated~\cite{Berninger2007a,Deachapunya2007a}.

Here we extend the operation of our deflectometer to the classical
Moir\'{e} mode with biomolecules and carbon nanotubes and we extend
the previous molecular measurement to an active sorting method for
molecular species that differ in $\alpha/m$.

\begin{figure}
\includegraphics[width=7cm]{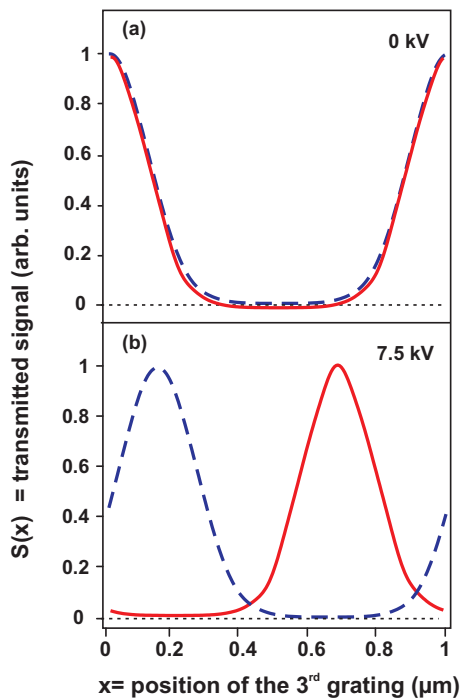}\\
\caption{Predicted fringe pattern for YGW and YWG tripeptides.
(a)shows the calculated density distribution after the third grating
without applying any voltage to the deflecting electrodes: The full
curve is the YWG and $--$ belongs to the YGW peptide. (b) indicates
that already at 7.5 kV both biomolecules can be separated and
therefore maximally enriched. The calculation takes also into
account the dispersive interaction of the molecules with the metal
gratings. The transmission function is periodic in x with over many
thousand lines, with a grating constant of 990 nm in this example.
\label{ywg_ygw_shift}}
\end{figure}

For a first illustration we discuss and simulate the relative
enrichment of a 50:50 mixture of the tripeptide
Tryptophan-Glycin-Tyrosin (YGW) and its isomer YWG which differ only
by the swapped position of Glycin and Tryptophan in the amino acid
sequence. Their masses are equal (m=460\,u) but their
susceptibilities $\chi\rm(YWG)=100\,\AA ^3$ and $\chi
\rm(YGW)=480\,\AA ^3$ differ by almost a factor of
five~\cite{Antoine2003a}. The susceptibility~\cite{Antoine2002a}
$\chi = \alpha + \langle \mu_{z}^{2}\rangle/(k_{B}T)$, includes the
orientation averaged square of the projection of the electric dipole
moment onto the direction of the external field $\langle \mu_{z}^2
\rangle$ and $T$ is the molecule temperature. With this definition,
the polarizability in Eq.~\ref{shift} may be replaced by $\chi$, if
the molecules also possess a permanent electric dipole moment.

For the two isomers the molecular fringe shifts will then differ by
a factor of five, if all other beam parameters are equal. Therefore,
when the three gratings are designed for maximum fringe contrast  in
the molecular beam close to the third grating, we may chose the
electric field such that one sort of peptide will be transmitted by
the deflectometer while its isomer will be blocked and deposited on
the third grating. The transmitted beam will then reveal a
significant enrichment of one particular isomer.

To quantify the sorting process we define the maximal
\emph{enrichment} of two mixed species $P_{1}$ and $P_{2}$ as:
\begin{equation}\label{enrichment}
  \eta=max_{|x}\{\tilde{S}_{P_{1}}(x)-\tilde{S}_{P_{2}}(x)\} ,
\end{equation}
where $\tilde{S}_{P_{i}}(x)= S(x)/[S_{max}(x)+S_{min}(x)]$ is the
normalized signal of the Moir\'{e} curve associated with the peptide
$P_{i}$ (see Fig.~2), and $x$ is the position of the third grating.

This definition is based on the fact that each isomer will form a
fringe pattern with its own intensity, fringe visibility and beam
shift in the external field gradient. Since the enrichmentis meant
to include only the effects of the sorting machine, the signals of
both species are normalized to their average beam fluxes.

The definition is chosen such that $\eta = 0$ for equal normalized
transmission of both species through the three-grating-arrangement,
and $\eta = 1$ if one species is blocked while the other is fully
transmitted.

For small polypeptides, the combination of a pulsed beam source with
a pulsed laser detection scheme, may allow us to select a mean
velocity of $v_y$= 340 m/s with a relative spread of $\Delta
v_y/v_y$ = 0.5\,\%. We now assume a grating separation of
$L=38.5\,$cm, a grating constant of 990\,nm, and a grating open
fraction of $f=0.2$, i.e gap openings of 200\,nm. Inserting all
these parameters we find a relative enrichment for YWG as high as
$\eta = 0.97$. The high expected degree of separation can also be
seen in Fig 2b. Here, the voltage has been optimized to
$(\mathbf{E\nabla} )E_{x}= 1.05\times 10^{13}$ \,V$^2$/m$^3$ in
order to maximize the transmitted content of this isomer. The
required field can be generated between two convex 5\,cm long
electrodes at a difference potential of U=7.5\,kV, and for a minimum
distance of 4\,mm.

Next to the sorting of biomolecules. The selection of carbon
nanotubes with a defined internal structure is a challenge that has
attracted great interest~\cite{Avouris2001}. Our deflectometer
proposal differs from earlier methods~\cite{Krupke2003a,Arnold2006a}
in that it is vacuum compatible and therefore better suited for a
certain class of technological applications. It also differs from a
recently patented suggestion for sorting free nanotube beams by
laser fields~\cite{Zhang2005} in that the use of microfabricated
gratings allows us to combine an uncollimated molecular beam with a
method of high spatial resolution.

\begin{figure}
\includegraphics[width=8.3cm]{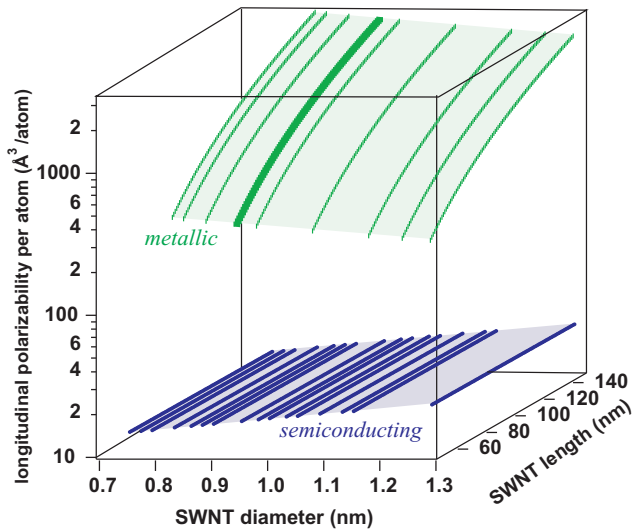}\\
\caption{Reduced longitudinal polarizability $\alpha_{\parallel}$
SWCNTs versus length and diameter. The two surfaces represent
$\alpha_{\parallel}$'s of metallic and semiconducting nanotubes of a
typical diameter~\cite{Hagen03} and possible length
distribution~\cite{heller2002}.\label{polari}}
\end{figure}

In the following we will assume that it is in principle possible
-- even though technically difficult at present -- to generate a
free molecular beam of single-wall carbon nanotubes (SWCNTs) with
an assumed length distribution between 50\,nm and 150\,nm, an
arbitrary mixture of chiralities and diameters between
0.7-1.3\,nm.

To simulate the Moir\'{e} fringes for these nanotubes we first need
to determine their $\alpha/m$ ratio. Their mass can be computed from
the number of carbon atoms per unit cell~\cite{Avouris2001}. The
static polarizability of nanotubes is extremely anisotropic and we
have to consider separately both the transverse and the longitudinal
value per carbon atom, i.e. the reduced polarizabilities. The
reduced transverse static polarizability of a carbon nanotube is
independent of its metallicity but it is proportional to its radius
$R$. For SWCNTs it can be approximated by $\alpha_{\perp red} \sim
1.3\AA^3$/atom\,\cite{benedict95}, a value very similar to that of
$C_{60}$ or medium-sized alkali clusters~\cite{knight85}.

The longitudinal polarizability of semiconducting tubes
$\alpha_{\parallel s}$ depends on their band gap energy
$E_{g}$\,\cite{benedict95} according to $\alpha _{\parallel s}
\propto \left(R/E_{g}^2\right)$. We use $\alpha_{\parallel s}
\approx 8.2 R^{2}+20.5$ for $R\geq 0.35$\,nm~\cite{Kozinsky06}. Even
for semiconducting SWCNTs the reduced longitudinal polarizability
thus exceeds already the transverse value by about a factor of ten
and the polarizability of medium-sized metal clusters by about a
factor of two~\cite{DeHeer1993a}.

This relation for $\alpha _{\parallel s}$ can't be applied to
metallic tubes because of their vanishing band gap, $E_{g}=0$. We
therefore approximate short metallic tubes of length $l$ by
perfectly conducting hollow cylinders ~\cite{Mayer2005a} and find
for their axial polarizability
\begin{equation}\label{pol_metallic}
    \alpha_{\parallel m} = \frac{l^3}{24(ln(l/R)-1)}\left(1+\frac{4/3 -
    ln(2)}{ln(l/R)-1}\right).
\end{equation}
This value exceeds that of equally long semiconducting tubes by a
factor between ten and one hundred. In Fig.~\ref{polari} we plot the
reduced polarizabilities for a range of different tube diameters and
lengths. The clear separation between metallic and semiconducting
tubes in this diagram indicates that mixtures of these species will
be well separable in a Moir\'{e}-deflection experiment.

The {\em reduced} longitudinal polarizability of semiconducting
tubes does not scale with the tube's length, since both their mass
and their polarizability grow linearly with it. The separation
process will therefore also work for nanotubes beyond the parameter
range of Fig.~\ref{polari}~\cite{Krupke2003a}.

With all masses and polarizabilities at hand, we now proceed to
simulate the Moir\'{e} fringe patterns.

\begin{figure}
\includegraphics[width=7cm]{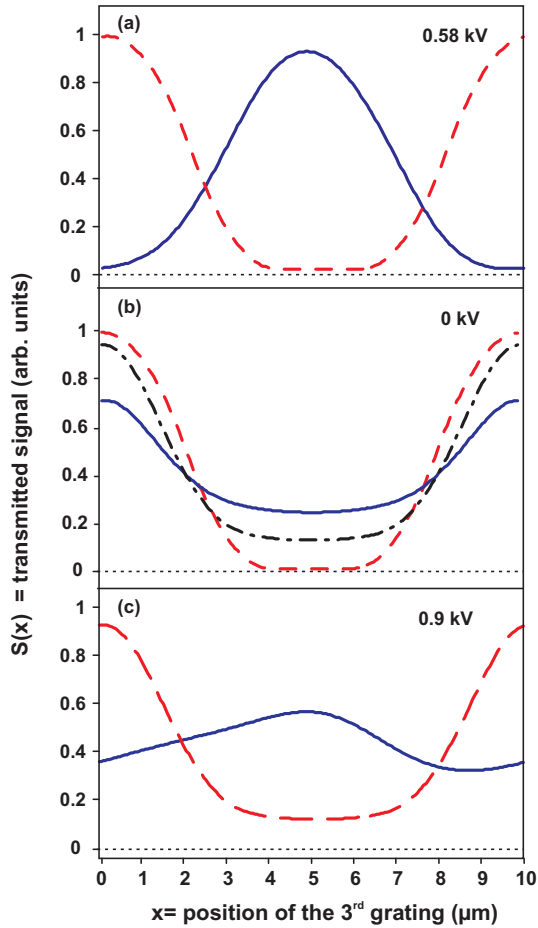}\\
\caption{Predicted fringe pattern for semiconducting (17,0) and
metallic (9,0) carbon nanotubes. (a) illustrates the ideal Moir\'{e}
case: The full curve is the (17,0) and $--$ belongs to the (9,0)
tube without Casimir-Polder (CP) and maximal aligned at 0.58~kV. (b)
shows the influence of the dispersive interaction between material
grating and nanotube: $--$ is the Moir\'{e} pattern. $-\cdot-$
includes the CP interaction for the (17,0) and the full curve for
the (9,0) tube at 0~kV~\cite{Bonin1997a} with maximum alignment,
i.e. without rotation. (c) is the complete analysis including CP and
full rotationl averaging: The full curve is the (9,0) and $--$ is
the (17,0) tube at 0.9~kV. \label{cntshift}}
\end{figure}

In Fig.~\ref{cntshift} we show the simulations for two 100\,nm long
semiconducting\,(17,0) and metallic\,(9,0) nanotubes flying at
100~m/s with a velocity spread of $\Delta v_y/v_y = 1\%$ through a
setup with metallic gratings separated by L=38.5 cm. The grating
period is now set to g = 10 $\rm \mu m$ and the open fraction is
again f=0.2, which would permit a fringe contrast of 100\% - for
small classical balls without polarizability.

The semiconducting tube is computed to have R=0.67\,nm,
m=$3.2\times10^{-22}$\,kg, $\alpha_{\perp} = 2.6\times
10^{4}\,{\AA}^3$ and $\alpha_{\parallel} = 3.8\times 10^5\,{\AA}^3$.
The metallic tube has R=0.36\,nm, m=$1.7\times 10^{-22}$\,kg,
$\alpha_{\perp} = 9.5\times 10^{3}\,{\AA}^3$ and $\alpha_{\parallel}
= 1.1\times 10^7\,{\AA}^3$. In the beginning we assume that all
nanotubes are maximally aligned with respect to the external
electric force field, i.e. along the x-axis. At a deflection field
of $(\mathbf{E} \mathbf{\nabla})E_{x} = 1.4\times 10^{12}\,V^2/m^3$,
the metallic tube's fringe shift of 5200\,nm would largely surpass
the 150\,nm shift of the semiconducting molecules. And one can
easily find a voltage that will enrich the metallic tubes in the
beam by shifting their fringe maxima until they fall onto the
openings of the third grating, while the semiconducting tubes will
be blocked by the grating bars. In this idealized picture the
enrichment could reach almost 100\% (Fig.~\ref{cntshift}A).

We now extend this simple model to include the attractive
Casimir-Polder (CP) potential  between the aligned molecules and
ideally conducting grating walls in the approximation of long
distances $r$:
\begin{equation}\label{cp}
   U(r) = -\frac{3 \hbar c }{8 \pi} \frac{\alpha}{r^{4}} ,
\end{equation}
~\cite{Casimir48}. The influence of the CP interaction is
demonstrated in Fig.~\ref{cntshift} (b). The fringe contrast is
reduced due to the deflection of the tubes in the grating's
potential. For this simulation metal gratings are assumed and a
larger enrichment can be maintained if the metal gratings are
replaced by dielectric materials or even by gratings made of
light~\cite{Nairz2001a, Gerlich2007a}.

We also have to consider that any nanotube beam in the forseeable
future will carry molecules in a highly excited rotational state.
Each orientation of the nanotube with respect to the external
electrode field is associated with a different fringe shift, since
the relative contributions by the transversal and longitudinal
polarizability depend on this orientation.

Fig.~\ref{cntshift}\,(c) shows an average of all Moir\'{e} curves
now including both the full rotational distribution
function~\cite{Bonin1997a} and the CP interaction. The expected
fringe visibility still amounts to 77\,\% for the semiconducting
(17,0) tubes and to 31\,\% for the metallic (9,0) ones. As can be
seen from Fig.~\ref{cntshift}\,(c) this will allow a significant
enrichment of the metallic tubes. The predicted value for the
enrichment reaches $\eta (17,0) = 0.4$ for the semiconducting tubes
and $\eta (9,0) = 0.6$ for the metallic ones. It is interesting to
see that our reasoning still holds generally for all other
chiralities. Metallic and semiconducting tubes will always be
separable with a good probability, because of the huge variation in
polarizabilities.

\begin{figure}
\includegraphics[width=6.5cm]{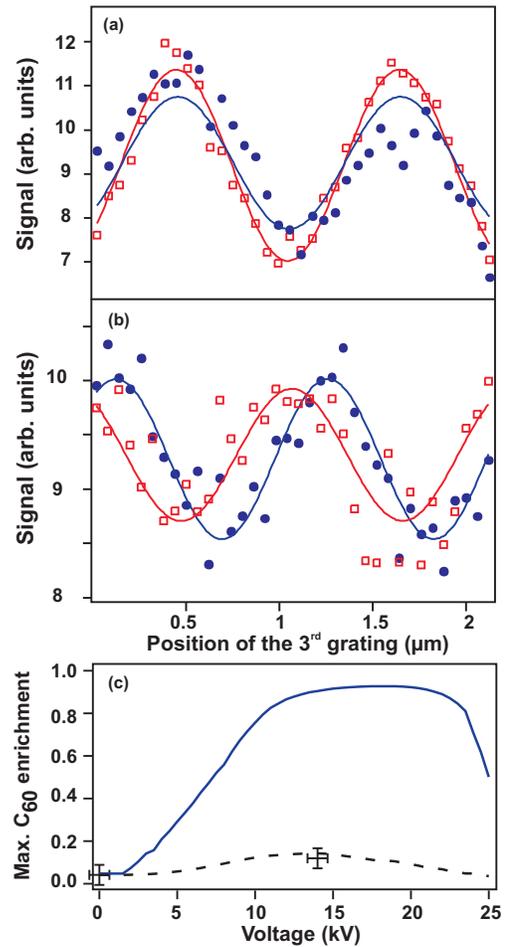}\\
\caption{A) Interference pattern without any voltage to the
electrodes. B) Separation of $C_{60}$ (circles) and $C_{70}$
(squares) at an electrode voltage of 14~kV. The phase shift
difference is $\delta =$171\,nm. Interference contrasts are
normalized to the same height. C) Comparison between expected
(dotted line) and observed maximal $C_{60}$ enrichment at 0 kV and
14 kV in the existing setup (crosses) with f=0.46 and $\Delta v/v =
15\%$. The potential for larger fullerene enrichment with an
optimized interferometer with g=990 nm, f=0.2 and $\Delta v/v = 1\%$
is indicated by the solid line. \label{c60shift}}
\end{figure}

To demonstrate the working principle of our three-grating sorting
machine we have performed experiments with the fullerens $C_{60}$
and $C_{70}$ in an existing Talbot-Lau interferometer with three
identical gold gratings with a period of g = 990 nm and an open
fraction of f=0.46. We detect the content of the different molecular
species using a quadrupole mass spectrometer (QMS Extrel, 2,000\,u).
The two fullerenes C$_{60}$ and C$_{70}$ differ in their mass by the
factor 7/6.

Their polarizability ratio was measured in a related experiment to
be $\alpha_{C_{70}}/\alpha_{C_{60}} =1.22$~\cite{Berninger2007a}.
The velocities in this mixture were $191$ m/s for $C_{60}$ and $184$
m/s for $C_{70}$, both with a velocity spread of 15\,\% from a
thermal source.  Fig.~\ref{c60shift}\,(a) shows the fringe contrast
of the two fullerenes without any voltage applied to the electrodes.
Even at $U=$0~kV we already observe a slight enrichment due to the
different fringe visibilities for $C_{60}$ and $C_{70}$. Applying a
voltage of 14 kV then results in the phase-shift difference shown in
Fig.~\ref{c60shift}\,(b). Fig.~\ref{c60shift}\,(c) plots the
measured and expected enrichments of $C_{60}$, which are in rather
good agreement.

The observed phase shift ratio $\Delta s (C_{70})/\Delta s(C_{60}) =
1.14$ fits well with our theoretical estimate (Eq.~\ref{shift})
 of $1.13$, including the statistical and
systematic error of 4\% in our experiment. For our experiment in
Fig.~\ref{c60shift}\,(b) we find a rather moderate $C_{60}$
enrichment of $\eta(C_{60})=0.08$. This is obviously not yet
optimized and it is interesting to discuss the factors that
influence it in the present and in future experiments.

Secondly, the fringe contrast is very sensitive to the van der Waals
interaction between the molecules and the grating walls. This
attractive potential modulates the fringe visibility and it does
this differently for different polarizabilities and molecule
velocities. This influence can be reduced by choosing a wider
grating period or by recurring to optical phase gratings, as
mentioned before~\cite{Brezger2003a}.

Thirdly, the Stark deflection itself is dispersive
(Eq.~\ref{shift}). A finite velocity spread leads to a reduction of
the interference contrast with increasing electric field. And while
the fringes in our present experiment would tend to wash out beyond
a deflection voltage of U=14\,kV, pulsed beams of
biomolecules~\cite{Marksteiner2006a} with $\Delta v_y/v_y \sim
0.1...1\%$ would be essentially free of such a restriction.

Fourthly, the polarizability ratio is rather small for the two
fullerene species. In contrast to that, $\alpha/m$ may vary by
$\sim500$\,\% for isomers of small polypeptides~\cite{Antoine2003a}
and by even a factor up to one hundred for carbon nanotubes of
different chirality~\cite{benedict95}. In this respect all future
experiments will be simpler compared to our present demonstration.

The very good quantitative agreement between our experiment and the
model expectations, shown in Fig.~\ref{c60shift}\,(c) proves that we
do understand the relevant processes in the present study. The solid
line in Fig.~\ref{c60shift}\,(c) shows the expected $C_{60}$
enrichment in an interferometer setup which is optimized for sorting
instead of quantum demonstrations.

Concluding, we have shown that $\alpha/m$-variations can be used to
sort neutral nanoparticles even in wide molecular beams. Our
simulations show that the relative enrichment may even get close to
100\% for biomolecular isomers and it will still be significant
($\sim 60\%$) for single-wall carbon nanotubes. The working
principle is illustrated by the enrichment of $C_{60}$ out of a
mixed molecular beam composed of $C_{60}$ and $C_{70}$ fullerenes.
The sorting scheme works in general for nanoparticles which can be
transferred into a free molecular beam and which differ in their
$\alpha$/m ratio.

\section*{Acknowledgments} This work has been supported by the
Austrian Science Funds (FWF) within the projects START177 and SFB
F1505. We acknowledge fruitful discussions with Klaus Hornberger. S.
D. acknowledges financial support by a Royal Thai government
scholarship.

\bibliographystyle{jacs}%{apsrev}

\end{document}